\documentclass[1p]{elsarticle}
\usepackage{amsmath,amssymb,amsfonts}
\usepackage{algorithmic}
\usepackage{graphicx}
\usepackage{subcaption}
\usepackage{textcomp}
\usepackage{eurosym}
\usepackage{arydshln} 

\usepackage{bm}
\usepackage{color}

\journal{Journal of Commodity Markets}

\begin{document}

\begin{frontmatter}

\title{Data-driven Calibration Sample Selection and Forecast Combination in Electricity Price Forecasting: An Application of the ARHNN Method}
\author{Tomasz Serafin\corref{cor1}}
\author{Weronika Nitka}

\affiliation{organization={Department of Operations Research and Business Intelligence,\\ 
           Wroc{\l}aw University of Science and Technology (WUST)},
            city={ Wroc{\l}aw},
            country={Poland}}

\cortext[cor1]{Corresponding author: Tomasz Serafin (e-mail: t.serafin@pwr.edu.pl).}

\begin{abstract}
Calibration sample selection and forecast combination are two simple yet powerful tools used in forecasting. They can be combined with a variety of models to significantly improve prediction accuracy, at the same time offering easy implementation and low computational complexity. While their effectiveness has been repeatedly confirmed in prior scientific literature, the topic is still underexplored in the field of electricity price forecasting.
In this research article we apply the Autoregressive Hybrid Nearest Neighbors (ARHNN) method to three long-term time series describing the German, Spanish and New England electricity markets. We show that it outperforms popular literature benchmarks in terms of forecast accuracy by up to 10\%. We also propose two simplified variants of the method, granting a vast decrease in computation time with only minor loss of prediction accuracy. Finally, we compare the forecasts' performance in a battery storage system trading case study. We find that using a forecast-driven strategy can achieve up to 80\% of theoretical maximum profits while trading, demonstrating business value in practical applications.
\end{abstract}


\begin{highlights}
\item This article describes different variants of ARHNN, a sample selection method.
\item The methods are applied to three diverse datasets from electricity markets.
\item Methods are evaluated in terms of statistical accuracy and business value.
\item ARHNN noticeably outperforms all benchmarks in most testing conditions.
\item Best financial results are not achieved by most statistically accurate forecasts.
\end{highlights}

\end{frontmatter}

\section{Introduction}
\label{sec:introduction}

Design and evaluation of models suitable for electricity price forecasting (EPF) has been widely studied in the scientific literature -- for extensive reviews see \citet{weron_electricity_2014,hong_energy_2020}. However, the accuracy of predictions depends not only on the chosen forecasting method, but also on the data used in the estimation process. The latter factor has been discussed in the econometric literature 
\citep{tian_forecast_2014,pesaran_selection_2007}, but within the context of EPF it has not received major attention so far. Meanwhile, it has been shown that the choice of calibration samples has a significant effect on forecasting accuracy \citep{hubicka_note_2019}, especially when structural breaks are present in data \citep{marcos_shortterm_2020, nasiadka_calibration_2022}.

The most well-known technique related to varying calibration samples is forecast combination (or forecast averaging). It is typically used in lieu of model selection, in order to reduce risks related to model selection and specification \citep{timmermann_chapter_2006,wang_forecast_2022}. As first demonstrated by \citet{hubicka_note_2019}, it can also be used to combine outputs of a single model specification with different calibration data, rather than a variety of models. In EPF literature this frequently results in robust forecasts with higher accuracy than any single expert model \citep{nitka_combining_2023,berrisch_multivariate_2024}. 

The ARHNN method \citep{nitka_forecasting_2021} has been among the first research articles in the field of EPF evaluating the effectiveness of estimating model parameters on calibration samples with nonconsecutive observations. It has been shown that including $k$-nearest neighbors as part of a hybrid forecasting method improves forecasting accuracy compared to standard literature benchmarks. However, the study had a limited scope, with data from a single energy market and no discussion on individual elements of the hybrid procedure.

This research paper has two major objectives: firstly, to propose several improvements to the ARHNN algorithm, targeting its interpretability, computational load and forecasting performance. Secondly, to comprehensively evaluate the discussed methods under different market conditions. The first objective is addressed by proposing three major modifications to the original version of ARHNN (Sec. \ref{ssec:methods_original_arhnn}): 1) updating the underlying model specification with new exogenous variables (Sec. \ref{ssec:methods_new_variables}), 2) fixing the sample size $k$ rather than using forecast averaging (Sec. \ref{ssec:methods_fixed_k}); 3) using weighted least squares estimation in place of the boolean sample choice with $k$-NN algorithm (Sec. \ref{ssec:methods_wls}). The second objective is achieved by applying the method to three datasets from electricity markets with different characteristics and analyzing both statistical accuracy and the economic benefits from using forecasts.

The $k$-nearest neighbors algorithm is attractive for applications in rapidly changing conditions, as it does not require pretraining. It has found successful uses in EPF, both directly \citep{lora_electricity_2007,fan_application_2019} and as part of hybrid models \citep{zhang_composite_2016}. Weighted least squares estimation has been used in EPF, to directly estimate forecasts \citep{reddy_dayahead_2016,hurd_modeling_2020} or to combine them \citep{bahman_longterm_2025}. Time-dependent weighted least squares has been used to correct for structural breaks in crude oil price forecasting \citep{wang_forecasting_2023}. The ARHNN method, originally proposed by \citet{nitka_forecasting_2021}, has been applied to forecast electricity prices by \citet{poggi_electricity_2023}. Our proposed modifications, while basing on well-established methods, have not been heretofore applied in electricity price modeling. Hence, we expand the forecasting literature by introducing and evaluating effective methods basing on calibration sample selection.

This article is structured as follows. Section \ref{sec:data} describes the data used in this research and its preparation. Section \ref{sec:methods} briefly describes the original ARHNN method, and subsequently introduces new modifications. Section \ref{sec:results} compares and evaluates the performance of modified ARHNN procedures in terms of forecast errors and a simulated practical case study. Finally, Section \ref{sec:conclusions} concludes and summarized the findings.

\section{Data}
\label{sec:data}

In this research article we display the performance of the ARHNN method when forecasting day-ahead electricity prices in three electricity markets with different characteristics. Firstly, the German EPEX SPOT market is an example of a temperate climate, with high penetration of renewables (from 33\,\% in 2015 to 75\,\% in 2024  \citep{burger_energycharts_2024}) and brown coal in the generation mix. The prices frequently exhibit drops into negative values. The second dataset describes the OMIE market in Spain. While it has a similarly high renewable share as Germany, it differs with regards to conventional generation technologies used. The most prevalent ones are natural gas and nuclear power. The same technologies dominate the generation mix of the third considered electricity market -- the ISO New England trading hub. However, compared to the European markets, the renewable energy shares are relatively low, making up only ca. 10\,\% of the total yearly generation \citep{isonewengland_iso_2025}.

The datasets span multiple years, allowing to test the methods' performance both before and during the 2022-2024 energy crisis. The European data includes electricity spot prices ($P$) from January 1, 2017 until December 21, 2024. The data sources are: \citep{entsoe_entsoe_2025}(EPEX-SPOT, OMIE) and \citep{isonewengland_iso_2025} (ISO-NE). Additionally, the data includes exogenous variables: the TSO forecasts of total system load ($\hat{L}$), wind energy generation ($\hat{W}$) and solar energy generation ($\hat{S}$). These variables are used to calculate forecasted residual load ($\hat{RL} = \hat{L} - (\hat{W} + \hat{S})$), used to forecast prices. They are supplemented by fuel price indices for coal ($C$) and natural gas ($G$), as well as EU emission allowance prices ($EUA$) (data: \citet{eia_opendata_2025}). Within the data, 728 days are reserved for the initial training and validation windows respectively, leaving four years (2021-2024) as the testing period. In the ISO-NE dataset, due to data availability, forecasts of renewable generation are replaced with one-day-ahead temperature forecast ($\hat{T}$) (data: \citet{zippenfenig_openmeteocom_2024}). All data are normalized before being subjected to further calculations.

All three markets exhibit high volatility, with the ISO-NE being the most consistent across the years. In the European markets, the range of values is increased by approximately an order of magnitude in years 2022-2024 compared to the previous average. The EPEX SPOT market experiences negative electricity prices very frequently compared to the other two datasets, with the OMIE prices capped at 0 EUR for the majority of the analyzed period, and the ISO-NE prices typically not dropping below approx. 20 USD. While the average price level clearly rises during the electricity price crisis in the ISO-NE market, the magnitude of outliers is not significantly larger than in earlier years. These major structural changes indicate not only a need for appropriate calibration sample selection, but also make an aggregated display of the results less informative. To that end, all results are presented with the test period split into calendar years.

In this article, we use the following nomenclature:
\begin{itemize}
    \item test set: last consecutive 6 (or 5) calendar years of the dataset, where forecasts are produced and evaluated in a rolling window scheme (i.e. with horizon of one day and most recent training data of a constant length, simulating real-life conditions);
    \item training set: part of the dataset available to forecast a single observation from the test set;
    \item calibration window: part of the training set consisting of most recent, consecutive observations, used to estimate model parameters;
    \item validation window: part of the training set consisting of most recent, consecutive observations, used to select hyperparameters;
    \item calibration sample: part of the training set used to estimate model parameters (not necessarily consecutive).
\end{itemize}

    \begin{figure*}
        \centering
        \includegraphics[width=\textwidth]{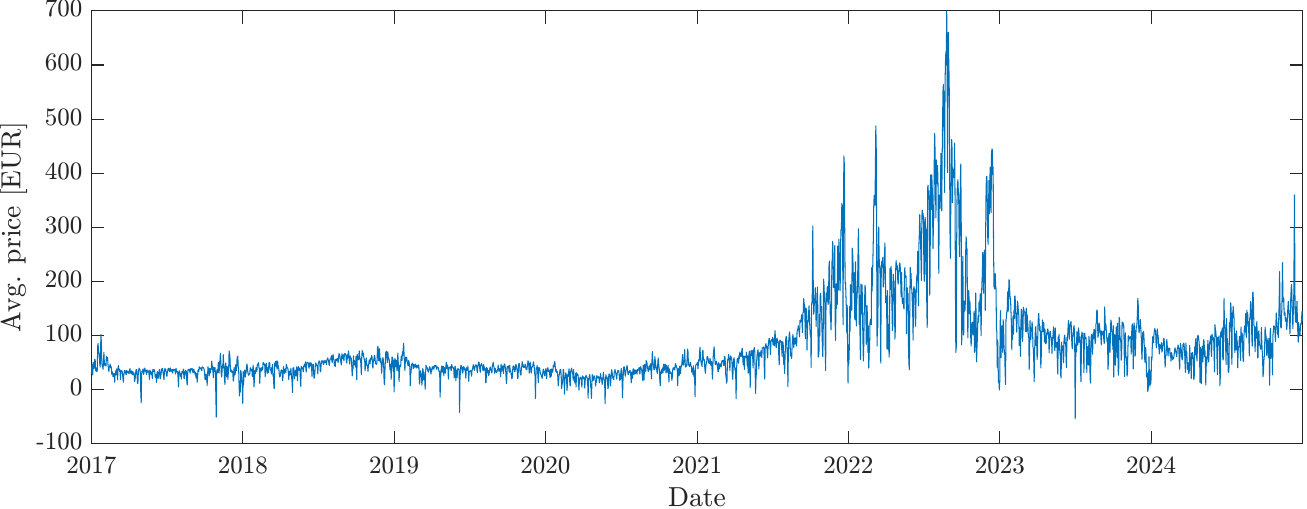}
        \caption{Daily price average, EPEX SPOT (01.01.2017--31.12.2024)}
        \label{fig:epex_data}
    \end{figure*}

    \begin{figure*}
        \centering
        \includegraphics[width=\textwidth]{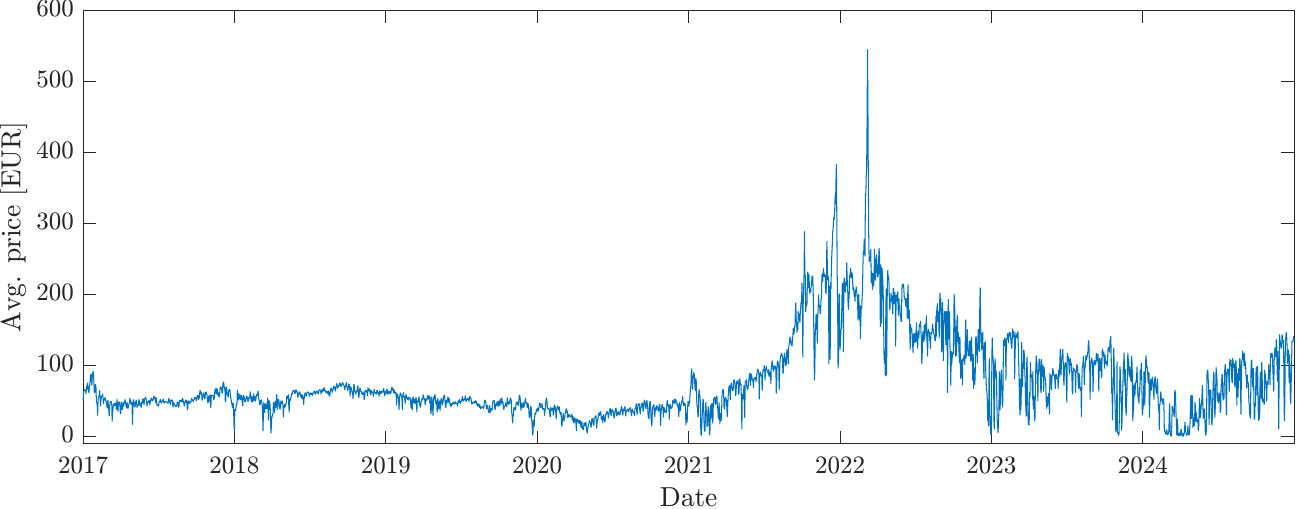}
        \caption{Daily price average, OMIE (01.01.2017--31.12.2024)}
        \label{fig:omie_data}
    \end{figure*}

    \begin{figure*}
        \centering
        \includegraphics[width=\textwidth]{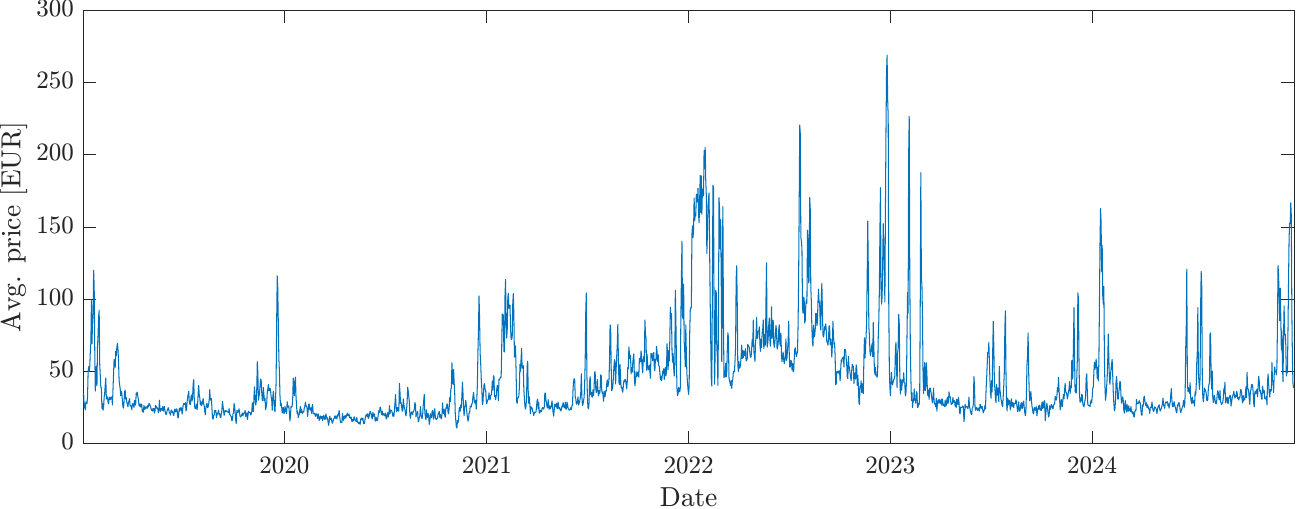}
        \caption{Daily price average, ISO-NE (01.01.2017--31.12.2024)}
        \label{fig:neiso_data}
    \end{figure*}

\section{Methods}
\label{sec:methods}

\subsection{ARHNN -- original version}
\label{ssec:methods_original_arhnn}

The main idea of the ARHNN method, as first proposed in \citet{nitka_forecasting_2021}, is to improve performance of a simple point forecasting model, such as a linear ARX model, by estimating its parameters on the most relevant training sample. If there are structural breaks in the data, they often have a negative impact on forecasting performance. In practice, it is very difficult to identify structural breaks, or regimes, in highly volatile data such as electricity prices. Different price behaviors depend both on long-term legal and geopolitical changes, and on rapidly evolving circumstances, e.g. weather conditions. Thus, choosing even a very short calibration window is not enough to ensure that the training data comes from a single regime. Moreover, for practical applications it is usually better to err on the side of using too much, rather than too little training data \citep{hubicka_note_2019}. 

The rationale behind ARHNN is to combine the robustness of a larger training dataset with quick adaptability of a small dataset. The relevancy is defined as the similarity of explanatory variables in the training sample compared to the current situation, i.e. "today" in a rolling window forecasting scheme. In other words, instead of estimating model parameters on a set of observations closest in time, it aims to choose the set closest in circumstances to the forecasted observation. By using $k$-NN to filter the original training sample, the estimated parameters should be true to the current data's regime, improving forecasting accuracy.

However, an optimal choice of hyperparameters is a non-trivial issue \citep{lago_forecasting_2021a, jedrzejewski_electricity_2022}. This is distinctly seen in context of time series, where suitable hyperparameters are likely to change over time. With ease of use and generality as major goals, ARHNN solves this issue by including a validation step in the procedure. Within the validation window, optimal values of $k$ are identified \textit{ex-post} for each observation. Finally, using this information, forecast averaging is leveraged to improve robustness and accuracy of the final forecast.

The steps of ARHNN are as follows:
\begin{enumerate}
    \item For all observations $P_T$ in the validation window and all considered values of $k$:
    \begin{enumerate}
        \item Construct vectors of exogenous variables $\Bar{X}_t$ for all observations in the calibration window and the target observation ($\Bar{X}_T$).
        \item Compute Euclidean distances $D_t$ between each $\Bar{X}_t$ and $\Bar{X}_T$.
        \item Choose $k$ closest observations (i.e. with smallest $D_t$) with $k$-NN as the calibration sample.
        \item Estimate forecast $\hat{P}_T^k$ using the calibration sample and compute its forecast error $\epsilon_T^k$.
    \end{enumerate}
    \item Choose the value of $k$ which minimizes $\epsilon_T$ and save it as $k_t$.
    \item For the target observation repeat steps 1a--1d for all values of $k_t$.
    \item Average all predictions from the previous step to obtain the final forecast.
\end{enumerate}

\subsection{Modification 1: alternate set of explanatory variables}
\label{ssec:methods_new_variables}

While the ARHNN method can be easily generalized to other applications, we demonstrate its performance for EPF. The original implementation followed a common literature approach of using autoregressive models with exogenous variables such as electricity demand and renewable energy generation. Following recent developments in this field, we note and incorporate several improvements to the model's specification:
\begin{itemize}
    \item solar energy generation is consistently equal to 0 during night hours, requiring special handling when implementing the algorithm using matrix operations;
    \item residual (net) load, defined as the difference between total demand and renewable energy generation, is highly relevant to price development \citep{denholm_overgeneration_2015,wang_enhancing_2017};
    \item external geopolitical circumstances, represented by fuel prices and EU carbon emission allowance prices, are a major driver of electricity prices in times of crisis.
\end{itemize}

Thus, the specification of ARX models used in the ARHNN procedure to forecast prices $P$ for day $d$, hour $h$ has been redefined as following:

\begin{equation}
\label{eq:ARX_DA}
\begin{split}
P_{d,h}  & = \underbrace{\alpha_hD_{d}}_{\text{Dummies}}+\underbrace{\sum_{p \in \{1,2,7\}} \beta_{h,p} P_{d-p,h}}_{\text{AR component}}
+\underbrace{\theta_{h,1}P_{d-1,24}}_{\text{Last known price}} + \\
& +\underbrace{\theta_{h,2}P_{d-1,min} +\theta_{h,3}P_{d-1,max}}_{\text{Yesterday's price range}} + \\
&+\underbrace{\theta_{h,4} \hat{RL}_{d,h} + \theta_{h,5} C_{d} + \theta_{h,6} G_{d} + \theta_{h,7} EUA_{d}}_{\text {Exogenous variables}} +  \varepsilon_{d,h},
\end{split}
\end{equation}
where variables follow notation introduced in Sec. \ref{sec:data}. Furthermore, the set of exogenous variables used to compute similarity has been changed to 
\begin{equation}
\left\lbrace P_{d-1,h}, P_{d-1,24}, P_{d-1,min}, P_{d-1,max}, \hat{RL}_{d,h}, C_{d}, G_{d}, EUA_{d}\right\rbrace.
\end{equation}
As in the original version, the most important explanatory variables are used in the $k$-NN step, omitting dummies and some lagged prices in order to reduce the problem's dimensionality. Henceforth, we denote this variant of the procedure as \textbf{ARHNN}. In Section \ref{sec:results} we compare the statistical performance of the original formulation of the model with the newly introduced variant.

While the Spanish market currently does not use coal power plants, their participation in the generation mix has been significant in the early parts of our dataset. Therefore, we have decided to include this variable in the model, making the above specification identical for the EPEX-SPOT and OMIE markets. For the ISO-NE market, due to different market characteristics (very small market share of coal), as well as lack of open access to renewable generation forecasts, the model specification is
\begin{equation}
\label{eq:ARX_DA_NEISO}
\begin{split}
P_{d,h}  & = \underbrace{\alpha_hD_{d}}_{\text{Dummies}}+\underbrace{\sum_{p \in \{1,2,7\}} \beta_{h,p} P_{d-p,h}}_{\text{AR component}}
+\underbrace{\theta_{h,1}P_{d-1,24}}_{\text{Last known price}} + \\
& +\underbrace{\theta_{h,2}P_{d-1,min} +\theta_{h,3}P_{d-1,max}}_{\text{Yesterday's price range}} + \\
&+\underbrace{\theta_{h,4} \hat{L}_{d,h} + \theta_{h,5} G_{d} + \theta_{h,6} \hat{T}_{d}}_{\text {Exogenous variables}} +  \varepsilon_{d,h},
\end{split}
\end{equation}
replacing the residual load and EU emission allowance price with total system load $\hat{L}$ and temperature forecast $\hat{T}$ respectively. Henry Hub gas spot prices are used to represent fuel price dynamics.

\subsection{Modification 2: averaging of predictions vs fixed $k$}
\label{ssec:methods_fixed_k}

As noted in Sec. \ref{ssec:methods_original_arhnn}, the main idea behind the ARHNN procedure is to filter the training data with $k$-NN. Additionally, to avoid the problem of arbitrarily selecting $k$ -- the most significant hyperparameter -- the method includes a validation step. For each observation in the validation window, the steps of filtering and estimating ARX parameters are repeated multiple times. We note that this addition offers significant benefits. Not only can ARHNN be applied nearly out-of-the-box on different datasets, typically its forecast accuracy is increased as well, thanks to combining multiple forecasts.

However, the increased complexity of the method induces drawbacks. Clearly, the most notable one is a significant increase in computational complexity. Instead of just a single instance of $k$-NN and ARX estimation, it is necessary to perform these two procedures for each reasonable value of $k$ and each observation in the validation window. This results in an approximately quadratical increase in computation time. While in the typical EPF context the entire training time is quite reasonable, counted in seconds for a single forecast, it could be problematic with larger datasets. Additionally, the additional calculations decrease the transparency and interpretability of the method, which can be an obstacle in practical usage and communication of the results.

Therefore, we aim to investigate whether the ARHNN procedure can be simplified by utilizing a fixed, user-selected value of $k$ instead of the validation step. The steps of this modified method, denoted henceforth \textbf{ARHNN($k$)}, are as follows:
\begin{enumerate}
    \item Construct vectors of exogenous variables $\Bar{X}_t$ for all observations in the calibration window and the target observation ($\Bar{X}_T$).
    \item Compute Euclidean distances $D_t$ between each $\Bar{X}_t$ and $\Bar{X}_T$.
    \item Choose $k$ closest observations (i.e. with smallest $D_t$) with $k$-NN as the calibration sample.
    \item Estimate forecast $\hat{P}_T$ using the calibration sample.
\end{enumerate}

For the practical application, we present forecast results for two arbitrary values of $k \in \left\lbrace 182, 364\right\rbrace$, representing respectively 25\% and 50\% of our chosen base calibration window size.

\subsection{Modification 3: replacing binary choice of data with weighted LS estimation}
\label{ssec:methods_wls}

The final modification investigates the feasibility of replacing the boolean $k$-NN selection of the calibration sample with a continuous measure of data points' relevancy. To that end, we replace the $k$-NN step with weighted least squares (WLS) estimation \citep{strutz_data_2015}. Commonly used to account for reliability or variance of observations in the training sample, we propose to use WLS to assign higher importance to relevant data points when estimating ARX parameters. The steps of the \textbf{WLS} procedure are: 
\begin{enumerate}
    \item Construct vectors of exogenous variables $\Bar{X}_t$ for all observations in the calibration window and the target observation ($\Bar{X}_T$).
    \item Compute Euclidean distances $D_t$ between each $\Bar{X}_t$ and $\Bar{X}_T$.
    \item Assign weights for each observation equal to the normalized inverse of the distance $\frac{1}{D_t \sum_{t\in 1,\ldots,T}{\frac{1}{D_T}}}$.
    \item Estimate forecast $\hat{P}_T$ using weighted least squares with the computed weights.
\end{enumerate}

Similarly to ARHNN($k$) (\ref{ssec:methods_fixed_k}), this procedure is not iterative. Since there are no hyperparameters to optimize and a fixed way to compute weights, only a single forecast is computed. This has a beneficial impact on computational complexity, which is comparable to ARHNN($k$). Another advantage of both of these approaches is their interpretability, allowing users to potentially gain insight from highly weighted (or selected) observations.

The downside of using the WLS variant over ARHNN lies in loss of robustness. In cases of very small distances between observations, the weights will end up extremely unbalanced. This is likely to have a negative impact on forecast accuracy due to introducing bias or numerical errors. Additionally, since forecast averaging is not included in this method, it cannot be used to stabilize results. However, we have not observed any notable numerical issues in our case study.

\section{Results}
\label{sec:results}

In this section we present the results of a case study where we apply different variants of ARHNN to forecast electricity prices. The data is described in detail in Sec. \ref{sec:data}. The evaluation is divided into two parts. Firstly, in Sec. \ref{ssec:results_accuracy}, we evaluate the forecast errors of each variant, comparing them to standard literature benchmarks. In order to focus attention on the benefits of calibration sample selection, we use a similar set of benchmarks as \citet{nitka_forecasting_2021}. All methods use an ARX model to forecast, with an identical specification presented in \eqref{eq:ARX_DA} and up to 728 days (two years) long calibration windows. The differences between the methods lie in two areas: observations used to estimate model parameters and use of forecast combinations.
In addition to four variants of ARHNN (\textbf{ARHNN}, see Sec. \ref{ssec:methods_new_variables}; \textbf{ARHNN(182)}, \textbf{ARHNN(364)}, see Sec. \ref{ssec:methods_fixed_k}; \textbf{WLS}, see Sec. \ref{ssec:methods_wls}), we consider three benchmark methods:
\begin{itemize}
    \item \textbf{Win(728)}, basic ARX forecast from a 728 day calibration window,
    \item \textbf{Avg(6)}, average of ARX forecasts from six hand-picked calibration windows with lengths $\left\lbrace56, 84, 112\right.$, $\left.714, 721, 728\right\rbrace$ days,
    \item \textbf{Avg(All)}, average of ARX forecasts from all 673 calibration windows with lengths between 56 and 728 days.
\end{itemize}
The benchmark methods use an identical model specification as in Equation \eqref{eq:ARX_DA} and ordinary least squares estimation.

\subsection{Forecast accuracy}
\label{ssec:results_accuracy}

This section compares the statistical accuracy of discussed methods. Models are compared using the root mean squared error ($RMSE = \sqrt{\frac{1}{n}\sum_{t=1,\ldots,n}{(\hat{P_t} - P_t)^2}}$) of all forecasts across the entire test period, i.e. calendar years 2021--2024. Due to large differences in price behaviors over time in all markets (see Sec. \ref{sec:data}), we report values for each year separately. 
Table \ref{tab:old_vs_new} presents the comparison of the ARHNN procedure based on the original (\textbf{ARHNN(old)}) and the newely introduced (\textbf{ARHNN}) specification of the ARX model for the EPEX dataset. As can be observed, the modification of the set of explanatory variables benefits the accuracy of forecasts for each year of the test period.

\begin{table}
\caption{RMSE of the ARHNN method based on the original and the new variant of the forecasting model per year in test period for the EPEX dataset. Top performers in each column are bolded.}
\label{tab:old_vs_new}
    \begin{tabular}{|l|cccc|}
\hline
Year &	2021	&	2022	&	2023	&	2024	\\
\hline
ARHNN &	\textbf{23.2669}	&	\textbf{41.8781}	&	\textbf{22.6366}	&	\textbf{34.0027}	\\
ARHNN(old) &	24.3673	&	43.0224	&	22.5859	&	34.9690	\\
\hline
    \end{tabular}
\end{table}

Tables \ref{tab:rmse_epex}--\ref{tab:rmse_neiso} display the root mean squared error of forecasts for the new ARHNN variant and benchmark approaches for the EPEX-SPOT, OMIE and ISO-NE datasets. It can be observed that among the benchmark methods, \textbf{Avg(6)} is typically the strongest performer, while \textbf{Win(728)} the weakest. This holds for all three datasets, although the differences are the most pronounced in the EPEX-SPOT market and the smallest in the ISO-NE market. This broadly corresponds to the variance of each dataset, with higher magnitude of spikes having a strong negative impact on forecast performance of the basic ARX-type models. Aside from one out of twelve cases (i.e. markets/years), all benchmark methods are outperformed by the full \textbf{ARHNN} method. Similarly to \textbf{Avg(6)}, it also tends to bring the most advantage in more volatile markets and time periods. 

Comparing these forecasts' performance of simplified methods: \textbf{ARHNN($k$)} and \textbf{WLS} it can be inferred that forecast averaging is the most important contributor to accuracy. Contrary to \textbf{ARHNN}, the modified versions struggle more frequently to outperform the \textbf{Avg(6)} benchmark. However, in most years, they offer at least a comparable performance, with \textbf{ARHNN(182)} typically achieving the most reliable results. Additionally, all simplified methods outperform the \textbf{Win(728)} benchmark by a large margin. This shows that a simple adjustment of calibration sample has a major effect on forecast accuracy, without a need to resort to more sophisticated methods.

While the full \textbf{ARHNN} method tends to outperform its competing forecasts in terms of accuracy, it requires a notable increase in computation time. On a personal computer (14th gen. Intel Core i7 processor, 64 GB RAM, MATLAB R2024A software), forecasts for the entire 4-year-long test set of the ISO-NE data have been generated within approximately 15 minutes (887.55~seconds). The same calculations for both \textbf{ARHNN($182$)} and \textbf{WLS} variants were more than 500 times faster, finishing within 1.33~s and 1.64~s respectively. While even the most computationally intensive variant is more than feasible to use in practice with similar datasets, it may be a factor if larger samples or many iterations of forecasts are needed by the users.

\begin{table}
\caption{RMSE of each method per year in test period, EPEX. Top performers in each column are bolded.}
\label{tab:rmse_epex}
    \begin{tabular}{|l|cccc|}
\hline
Year &	2021	&	2022	&	2023	&	2024	\\
\hline
Win(728) &	28.3809	&	50.5249	&	29.2686	&	36.7748	\\
Avg(6) &	23.7878	&	43.8228	&	23.8784	&	34.7704	\\
Avg(All) &	25.3391	&	45.0916	&	27.2313	&	34.5458	\\
\hdashline
ARHNN &	\textbf{23.2669}	&	\textbf{41.8781}	&	\textbf{22.6366}	&	\textbf{34.0027}	\\
ARHNN(182) &	23.4102	&	43.0263	&	22.8827	&	34.7438	\\
ARHNN(364) &	25.9670	&	45.2098	&	23.9814	&	34.9389	\\
WLS &	26.4706	&	46.4781	&	24.3678	&	34.2225	\\
\hline
    \end{tabular}
\end{table}

\begin{table}
\caption{RMSE of each method per year in test period, OMIE. Top performers in each column are bolded.}
\label{tab:rmse_omie}
    \begin{tabular}{|l|cccc|}
\hline
Year &	2021	&	2022	&	2023	&	2024	\\
\hline
Win(728)	&	19.8744	&	31.1978	&	19.2686	&	17.1868	\\
Avg(6) 	&	18.8146	&	29.4001	&	\textbf{17.1814}	&	15.0704	\\
Avg(All)	&	18.7743	&	30.3211	&	17.4397	&	15.3077	\\
\hdashline
ARHNN	&	\textbf{18.3266}	&	\textbf{28.8995}	&	17.3749	&	\textbf{14.5205}	\\
ARHNN(182)	&	18.4171	&	30.7583	&	18.4660	&	14.8463	\\
ARHNN(364)	&	18.9393	&	30.7775	&	17.9438	&	15.4087	\\
WLS	&	18.9828	&	30.2183	&	18.0680	&	15.5807	\\
\hline
    \end{tabular}

\end{table}

\begin{table}
\caption{RMSE of each method per year in test period, ISO-NE. Top performers in each column are bolded.}
\label{tab:rmse_neiso}
    \begin{tabular}{|l|cccc|}
\hline
Year &	2021	&	2022	&	2023	&	2024	\\
\hline
Win(728)	&	12.0558	&	22.8456	&	15.8806	&	14.7688	\\
Avg(6) 	&	11.7426	&	21.8928	&	14.2483	&	15.2164	\\
Avg(All)	&	11.7764	&	21.9939	&	15.1574	&	14.4581	\\
\hdashline
ARHNN	&	\textbf{11.4116}	&	\textbf{21.5267}	&	\textbf{12.7586}	&	\textbf{13.2361}	\\
ARHNN(182)	&	11.6112	&	21.5835	&	13.0617	&	13.7176	\\
ARHNN(364)	&	11.7778	&	22.0795	&	13.7752	&	13.9351	\\
WLS	&	11.7650	&	22.1744	&	13.9623	&	13.7189	\\
\hline
    \end{tabular}
\end{table}

\begin{figure}
\centering
\begin{minipage}{.4\textwidth}
    \centering
    \includegraphics[width=\linewidth]{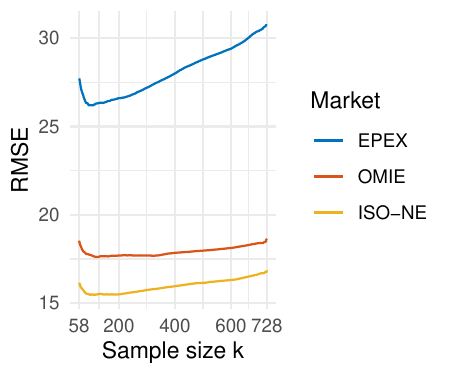}
    \captionof{figure}{RMSE of \textbf{ARHNN($k$)} across entire test period for all values of $k$, all datasets.}
    \label{fig:all_rmse_k}
\end{minipage}%
\hspace{0.5cm}
\begin{minipage}{.55\textwidth}
  \centering
    \includegraphics[width=\linewidth]{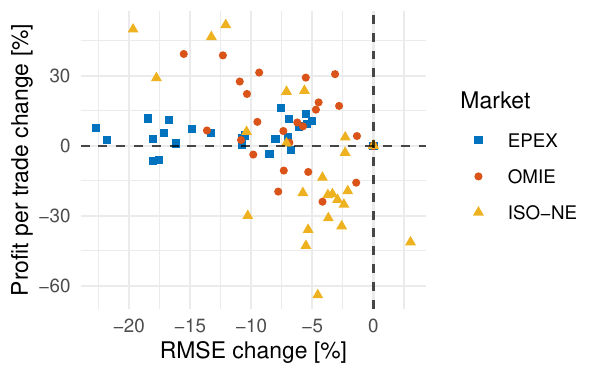}
    \captionof{figure}{Percentage change in profit per trade versus forecast error compared to \textbf{Win(728)}, per forecast and year; colors indicate market.}
    \label{fig:stat_econo_ppt}
\end{minipage}
\end{figure}

In order to clearly present the results in tabular form, two values of $k$ have been hand-picked for the \textbf{ARHNN($k$)} method. While the chosen values are easy to understand and interpret in a practical context, they necessarily cannot be an optimal choice for this hyperparameter. Therefore, we take a closer look at the impact of this hyperparameter's choice on forecast errors. Figure \ref{fig:all_rmse_k} presents the RMSE of \textbf{ARHNN($k$)} for each value of $k$ in the entire test period. When analyzed separately, the general shape of the curve is similar, but the minimum changes its location. For more volatile periods best $k$ tends to be lower (100 and less), while for periods of lower prices it tends to be between 200 and 300. These observations hold for all three markets.

\subsection{Economic value -- daily arbitrage with a battery storage system}
\label{ssec:results_econo}


\begin{table}
\caption{Economic measures of each method per year in test period, EPEX. Top performers in each column are bolded.}
\label{tab:econo_epex}
    \begin{tabular}{|l|cccc|}
    \hline
    \multicolumn{5}{|c|}{Total profit (in \euro{})} \\
\hline
Year	&	2021	&	2022	&	2023	&	2024	\\
\hline
Win(728)	&	5028	&	23200	&	7490	&	19613	\\
Avg(6) 	&	\textbf{5534}	&	24336	&	8048	&	20140	\\
Avg(All)	&	5371	&	24034	&	7614	&	20265	\\
ARHNN	&	5125	&	24439	&	7801	&	20553	\\
ARHNN(182)	&	5521	&	\textbf{24896}	&	7775	&	20090	\\
ARHNN(364)	&	5251	&	24439	&	7853	&	\textbf{20624}	\\
WLS	&	5257	&	23637	&	\textbf{8084}	&	20355	\\
Crystal ball & 7756 & 29759 & 10932 & 25659 \\
    \hline
    \hline
    \multicolumn{5}{|c|}{Profit per trade (in \euro{})} \\
\hline
Win(728)	&	45.3	&	67.8	&	25.9	&	64.1	\\
Avg(6) 	&	\textbf{45.7}	&	71.6	&	\textbf{28.9}	&	70.2	\\
Avg(All)	&	45.5	&	70.1	&	26.9	&	69.4	\\
ARHNN	&	42.4	&	71.5	&	27.9	&	\textbf{74.5}	\\
ARHNN(182)	&	42.5	&	\textbf{72.8}	&	26.5	&	72.8	\\
ARHNN(364)	&	43.8	&	70.8	&	26.7	&	70.9	\\
WLS	&	44.6	&	69.9	&	28.8	&	71.4	\\
Crystal ball & 57.9 & 97.9 & 49.2 & 92.0 \\
    \hline
    \hline
    \multicolumn{5}{|c|}{Sharpe ratio} \\
\hline
Win(728)	&	0.84	&	0.82	&	0.46	&	0.71	\\
Avg(6) 	&	0.88	&	0.87	&	\textbf{0.51}	&	0.77	\\
Avg(All)	&	\textbf{0.90}	&	0.85	&	0.48	&	0.77	\\
ARHNN	&	0.81	&	0.88	&	0.5	&	\textbf{0.81}	\\
ARHNN(182)	&	0.87	&	\textbf{0.89}	&	0.47	&	\textbf{0.81}	\\
ARHNN(364)	&	0.85	&	0.87	&	0.48	&	0.79	\\
WLS	&	0.87	&	0.85	&	\textbf{0.51}	&	0.8	\\
Crystal ball & 1.24 & 1.26 & 0.91 & 1.03 \\
\hline
    \end{tabular}
\end{table}


\begin{table}
\caption{Economic measures of each method per year in test period, OMIE. Top performers in each column are bolded.}
\label{tab:econo_omie}
    \begin{tabular}{|l|cccc|}
    \hline
    \multicolumn{5}{|c|}{Total profit (in \euro{})} \\
\hline
Year	&	2021	&	2022	&	2023	&	2024	\\
\hline 
Win(728)	&	439	&	3347	&	3039	&	2748	\\
Avg(6) 	&	605	&	3789	&	2977	&	\textbf{3794}	\\
Avg(All)	&	\textbf{681}	&	\textbf{4422}	&	3084	&	3520	\\
ARHNN	&	548	&	4154	&	2882	&	3666	\\
ARHNN(182)	&	648	&	3668	&	2312	&	2956	\\
ARHNN(364)	&	609	&	4113	&	3035	&	3152	\\
WLS	&	626	&	4404	&	\textbf{3177}	&	3422	\\
Crystal ball & 1788 & 7176 & 5512 & 6024 \\

    \hline
    \hline
    \multicolumn{5}{|c|}{Profit per trade (in \euro{})} \\
\hline
Win(728)	&	21.97	&	21.46	&	15.2	&	12.96	\\
Avg(6) 	&	19.51	&	23.25	&	15.58	&	17.98	\\
Avg(All)	&	\textbf{28.39}	&	25.13	&	\textbf{16.76}	&	16.53	\\
ARHNN	&	17.66	&	22.82	&	14.63	&	\textbf{18.06}	\\
ARHNN(182)	&	19.64	&	18.07	&	11.56	&	13.81	\\
ARHNN(364)	&	25.37	&	22.36	&	15.41	&	15.84	\\
WLS	&	26.06	&	\textbf{28.05}	&	16.72	&	17.03	\\
Crystal ball & 38.0 & 43.8 & 34.2 & 31.9 \\
    \hline
    \hline
    \multicolumn{5}{|c|}{Sharpe ratio} \\
\hline
Win(728)	&	0.51	&	0.51	&	0.47	&	0.39	\\
Avg(6) 	&	0.4	&	0.54	&	0.48	&	\textbf{0.59}	\\
Avg(All)	&	\textbf{0.60}	&	0.58	&	0.51	&	0.52	\\
ARHNN	&	0.4	&	0.52	&	0.44	&	0.58	\\
ARHNN(182)	&	0.45	&	0.42	&	0.35	&	0.43	\\
ARHNN(364)	&	0.57	&	0.51	&	0.47	&	0.49	\\
WLS	&	0.52	&	\textbf{0.62}	&	\textbf{0.52}	&	0.53	\\
Crystal ball & 0.85 & 1.12 & 1.37 & 1.38 \\
\hline
    \end{tabular}
\end{table}


\begin{table}
\caption{Economic measures of each method per year in test period, ISO-NE. Top performers in each column are bolded.}
\label{tab:econo_neiso}
    \begin{tabular}{|l|cccc|}
    \hline
    \multicolumn{5}{|c|}{Total profit (in \$)} \\
\hline
Year	&	2021	&	2022	&	2023	&	2024	\\
\hline 
Win(728)	&	130	&	1117	&	295	&	882	\\
Avg(6) 	&	146	&	\textbf{1654}	&	218	&	706	\\
Avg(All)	&	180	&	1436	&	148	&	872	\\
ARHNN	&	131	&	1351	&	394	&	1062	\\
ARHNN(182)	&	\textbf{205}	&	1077	&	360	&	932	\\
ARHNN(364)	&	173	&	1266	&	385	&	1040	\\
WLS	&	125	&	1178	&	\textbf{399}	&	\textbf{1086}	\\
Crystal ball & 640 & 3512 & 1103 & 1597 \\
    \hline
    \hline
    \multicolumn{5}{|c|}{Profit per trade (in \$)} \\
\hline
Win(728)	&	18.53	&	\textbf{31.91}	&	16.41	&	40.07	\\
Avg(6) 	&	12.15	&	27.57	&	11.49	&	23.53	\\
Avg(All)	&	17.99	&	25.2	&	5.92	&	32.31	\\
ARHNN	&	11.87	&	25.48	&	24.6	&	42.48	\\
ARHNN(182)	&	12.8	&	18.25	&	21.18	&	40.52	\\
ARHNN(364)	&	\textbf{19.21}	&	25.32	&	24.07	&	\textbf{49.54}	\\
WLS	&	13.87	&	24.54	&	\textbf{24.91}	&	49.37	\\
Crystal ball & 25.6 & 38.2 & 36.8 & 69.4 \\

    \hline
    \hline
    \multicolumn{5}{|c|}{Sharpe ratio} \\
\hline
Win(728)	&	0.48	&	\textbf{0.59}	&	0.3	&	0.52	\\
Avg(6) 	&	0.29	&	0.6	&	0.21	&	0.32	\\
Avg(All)	&	0.44	&	0.53	&	0.12	&	0.43	\\
ARHNN	&	0.32	&	0.53	&	\textbf{0.48}	&	0.57	\\
ARHNN(182)	&	0.36	&	0.36	&	0.42	&	0.53	\\
ARHNN(364)	&	\textbf{0.54}	&	0.51	&	0.46	&	0.63	\\
WLS	&	0.32	&	0.48	&	0.45	&	\textbf{0.67}	\\
Crystal ball & 1.19 & 1.06 & 0.97 & 1.14 \\

\hline
    \end{tabular}
\end{table}

This evaluation follows the assumptions and framework proposed by \citet{serafin_loss_2025}. Day-ahead electricity price forecasts are used to select a pair of hours, ($h_1$, $h_2$), that maximize the value of a price spread $\Delta \hat{P}_{d,h_1,h_2}$:
\begin{equation}
    \Delta \hat{P}_{d,h_1,h_2} = 0.9 \hat{P}_{d,h_2} - \frac{1}{0.9} \hat{P}_{d,h_1},
\end{equation}
where $\hat{P}_{d,h}$ is the electricity price forecast for day $d$ and hour $h$ and coefficients represent the battery charging/discharging efficiency of 90\%. Similarly to \citet{serafin_loss_2025}, the strategy uses a predefined threshold level $T$, to filter most profitable trading opportunities. More precisely, price taker orders are submitted in the day-ahead market only if the estimated profit exceeds the threshold:
\begin{equation}
    \Delta \hat{P}_{d,h_1,h_2} \geq T.
\end{equation}
Here, we assume $T=50$ \euro{} and a per-cycle battery cost equal to $T$ although the realistic estimation of current BESS operation costs is closer to 100 \euro{}, see \citet{serafin_loss_2025}).

Three measures for the economic evaluation of forecasts are considered in the paper: total profit, profit per trade and the Sharpe ratio. The total profit (TP) is given by the following formula:
\begin{equation}
    TP =\sum_{d=1,\ldots,n}p_d,
\end{equation}
where $p_d$ is the daily profit from trading on day $d$. As the number of trading days may vary between models (and TP may favor quantity over quality of trades), profit per trade (PPT) measures the average quality of a single trade:
\begin{equation}
   PPT = \frac{TP}{\mathrm{No. Trades}},
\end{equation}
where the denominator represents the number of days when the model traded. Lastly, we consider the Sharpe ratio (SR) for the evaluation of risk-adjusted profits:
\begin{equation}
   SR = \frac{PPT}{\sigma},
\end{equation}
where $\sigma$ is the standard deviation of non-zero profits (excluding days in which a certain model did not trade).

Tables \ref{tab:econo_epex}--\ref{tab:econo_neiso} report results of the economic evaluation for three considered datasets. It can be seen that unlike in the case of error metrics, it is difficult to pinpoint the consistent top-performer. The best performing models differ across datasets, years and metrics. 

\subsubsection{Total profit}
For the German day-ahead market, in the calm year 2021, strategy based on forecasts from model \textbf{Avg(6)} is able to achieve the highest total profit, very closely followed by \textbf{ARHNN(182)}. The latter model generates highest revenue in the most profitable year 2022, outperforming \textbf{ARHNN} and \textbf{ARHNN(364)} by ca. 400 \euro{}. In 2023 and 2024 again the sample selection-based methods (\textbf{WLS} in 2023 and \textbf{ARHNN(364)} in 2024) produce forecasts maximizing total profits from day-ahead trading.

For the Spanish market, the literature benchmarks of \citet{hubicka_note_2019}, namely \textbf{Avg(6)} and \textbf{Avg(All)} perform best in 3 out of 4 evaluated years (2021, 2022 and 2024). The WLS manages to outperform all other approaches in year 2024. 

For the ISO-NE dataset, \textbf{ARHNN(182)} is again the best performer for year 2021. \textbf{Avg(6)} outperforms other approaches in the most profitable year 2022 and for the rest of the evaluation period (years 2023 and 2024), the \textbf{WLS} model generates highest revenues. Its worth noting that the magnitude of revenues for OMIE and ISO-NE is noticeably lower than for the German market.

\subsubsection{Profit per trade}
For the EPEX dataset, the \textbf{Avg(6)} benchmark is the top performer for calm years 2021 and 2023. In years 2022 and 2024, forecasts from models \textbf{ARHNN(182)} and \textbf{ARHNN} respectively, allow for the best quality trades.

For OMIE, the \textbf{Avg(All)} model outperforms every other approach for years 2021 and 2023, while sample selection-based models, i.e. \textbf{WLS} and \textbf{ARHNN}, perform best in years 2022 and 2024, respectively. 

\textbf{ARHNN(364)} produces forecasts characterized by the highest profit per trade for years 2021 and 2024 for the ISO-NE dataset. The \textbf{WLS} approach places second in 2024 and is the top performer in year 2023. Interestingly, in year 2022, the simplest approach -- namely \textbf{Win(728)} -- manages to achieve the best profit per trade, although the total profit is lower than for other methods.

\subsubsection{Sharpe ratio}
For the risk adjusted profits, ranking of the models often corresponds to the one for the profit per trade. For EPEX, \textbf{Avg(All)} is the best performer for year 2021, \textbf{ARHNN(182)} for year 2022, \textbf{Avg(6)} and \textbf{WLS} tie in year 2023 and in 2024 \textbf{ARHNN} and \textbf{ARHNN(182)} outperform all other approaches. 

For the OMIE dataset, the same models -- \textbf{Avg(All)} and \textbf{WLS} -- are top performers for years 2021 and 2022 in terms of profit per trade and the Sharpe ratio. \textbf{WLS} again performs best in 2023 and for 2024 best Sharpe ratio is achieved for forecasts from the \textbf{Avg(6)} benchmark.

For ISO-NE, similarly to OMIE, the same models rank best for Sharpe ratio and profit per trade in years 2021 and 2022 (\textbf{Win(728)} and \textbf{ARHNN(364)}). \textbf{ARHNN} performs best in 2023 while for year 2024 the \textbf{WLS} model is the best performer.     

\subsection{Statistical vs economic value -- discussion}
\label{ssec:results_econo_disc}

As \citet{yardley_errors_econo} claim, the literature generally reveals a disagreement between traditional error metrics and economic measures of performance. While several papers in the econometric literature support this statement \citep{cenesizoglu_value,caldeira_yield}, this aspect has been only recently addressed in the EPF literature \citep{nitka_combining_2023}. 

\begin{figure}
\centering
\begin{minipage}{.45\textwidth}
    \centering
    \includegraphics[width=\linewidth]{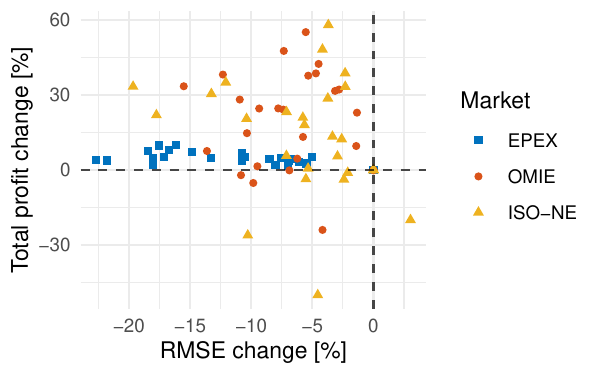}
    \captionof{figure}{Percentage change in total profit versus forecast error compared to \textbf{Win(728)}, per forecast and year.}
    \label{fig:stat_econo_profit}
\end{minipage}%
\hspace{0.5cm}
\begin{minipage}{.45\textwidth}
  \centering
    \includegraphics[width=\linewidth]{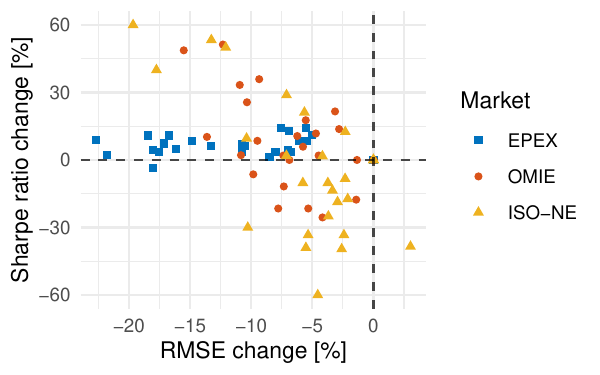}
    \captionof{figure}{Percentage change in Sharpe ratio versus forecast error compared to \textbf{Win(728)}, per forecast and year.}
    \label{fig:stat_econo_sharpe}
\end{minipage}
\end{figure}

Since this paper involves both statistical and economic evaluation of day-ahead electricity price forecasts, the relationship of both approaches is discussed here. Firstly, it can be noticed that although the \textbf{ARHNN} model outperforms all other approaches in terms of RMSE for nearly all years and datasets, the performance does not translate to superior economic results with respect to any of the considered metrics. In order to entangle the relationship between the economic and statistical performance of models, the following exercise was performed. First, the worst performing benchmark (in terms of RMSE), \textbf{Win(728)}, was selected as the reference point. Second, a relative improvement in the statistical and economic measures over the benchmark was calculated for each year, model and dataset. Figures \ref{fig:stat_econo_profit}--\ref{fig:stat_econo_sharpe} present scatterplots of the aforementioned differences for the total profit and the Sharpe ratio (results are very similar for the profit per trade and the Sharpe ratio, thus the former is not reported). 

A few interesting observations can be drawn from Figures \ref{fig:stat_econo_profit}--\ref{fig:stat_econo_sharpe}. First, in case of the EPEX dataset, both for profit per trade and Sharpe ratio there seems to be no correlation between the statistical and economic improvement in performance over the \textbf{Win(728)} benchmark. Second, for OMIE and ISO-NE markets, even though the correlation of economic and statistical performance is more visible, only the best performing methods (RMSE-measured) managed to improve economic metrics over the selected benchmark. In several cases a notable reduction of RMSE actually corresponded with an underperformance with respect to total trading profits and the Sharpe ratio. This is especially the case for the ISO-NE market. This demonstrates that in practice, business value of forecasts needs to be analyzed alongside their statistical properties.

\section{Conclusions}
\label{sec:conclusions}

This article aims to propose and evaluate sample selection methods for EPF. We describe the ARHNN method and introduce its three modified variants. We apply them to three large datasets from three electricity markets with varying characteristics and evaluate their performance in terms of statistical accuracy and business value. The test period covers both highly turbulent prices and calmer periods.

The results show that the full ARHNN method is highly successful in forecasting in nearly every conditions, noticeably outperforming all benchmarks and simplified methods in most test cases. The simplified methods are usually competitive with more sophisticated benchmarks, heavily outperforming the simplest ARX forecast. However, as previously demonstrated in literature, forecast combination methods are very robust and difficult to significantly outperform. This can be seen in the presented results as well. Additionally, among the simplified methods, there is no single variant which universally performs well. On the contrary, the best choice out of sample selection methods without averaging changes across datasets and in time. This shows that within the presented approaches, there exists a trade-off between forecasting accuracy and computational complexity.

Although ARHNN has a consistently superior performance in terms of statistical accuracy, the business value case study gives less clear-cut results. While there exists a slight correlation between improvement of profits and forecast errors, it is not enough to determine models with the best business value. Oftentimes, the forecast best in statistical terms is outperformed in the arbitrage trading case study by much less accurate models. Additionally, the precise ranking of economic results within a single market varies significantly from year to year. However, all forecast-driven strategies achieve high profits in relation to maximum potential "crystal ball" gains.

We believe that the models presented in this paper offer high value to industry practitioners, due to their relatively easy implementation and straightforward interpretability.

\section*{Acknowledgments}
This work was supported by National Science Center (NCN, Poland) through MAESTRO grant No. 2018/30/A/HS4/00444 (to W.N.) and SONATA BIS grant \\ No. 2019/34/E/HS4/00060 (to T.S.).

The authors would like to extend thanks to prof. Micha{\l} Rubaszek, SGH Warsaw School of Economics, Warsaw, Poland, for sparking a discussion regarding simplified methods presented in this research paper.

\bibliographystyle{elsarticle-harv}
\bibliography{Nitka_Serafin_data_driven}

\end{document}